\documentclass[runningheads]{llncs}

\usepackage{hyperref}
\usepackage[utf8]{inputenc} 
\usepackage{amssymb}
\usepackage{listings} 
\lstdefinestyle{Muli}{
	language=Java,
	morekeywords={free,fail,solve,getAllSolutions,getAllSolutionsEx,getOneSolution,getOneSolutionEx,search},
	captionpos=b,
	tabsize=2,
	belowcaptionskip={10\p@},
	abovecaptionskip={0\p@},
	showstringspaces=false,
	basicstyle=\ttfamily 
}
\usepackage{cleveref} 
\usepackage{microtype}

\usepackage{cleveref}
\usepackage{graphicx} 
\usepackage[caption=false]{subfig} 
\usepackage{tikz} 
\usetikzlibrary{calc}
\usetikzlibrary{positioning} 
\usetikzlibrary{arrows} 
\usetikzlibrary{shapes.geometric} 
\usetikzlibrary{matrix,fit} 
\usepackage{tikz-qtree}
\usepackage{tikz-uml}
\usepackage{erciscolors}

\usepackage{booktabs} 

\usepackage{pgfplots}
\def\addlegendimage{\csname pgfplots@addlegendimage\endcsname}
\pgfkeys{/pgfplots/number in legend/.style={%
		/pgfplots/legend image code/.code={%
			\node [anchor=east]at (0.7,-0.0225){#1}; 
		},%
	},
}

\begin{document}
	\renewcommand{\thelstlisting}{\arabic{lstlisting}} 

	\title{Reference Type Logic Variables in Constraint-logic Object-oriented Programming}
	\titlerunning{Reference Type Logic Variables in Constraint-logic OO Programming}

	\author{Jan C. Dageförde\orcidID{0000-0001-9141-7968} 
		}
	\authorrunning{J. C. Dageförde}
	\institute{ERCIS, Leonardo-Campus 3, D-48149 Münster, Germany\\
		\email{dagefoerde@uni-muenster.de}}

	\maketitle              

	\begin{abstract}
	Constraint-logic object-oriented programming, for example using Muli, facilitates the integrated development of
	business software that occasionally involves finding solutions to constraint-logic problems.
	The availability of object-oriented features calls for the option to use objects as logic variables as well, as opposed to being limited to primitive type logic variables. 
	The present work contributes a concept for reference type logic variables in constraint-logic object-oriented programming
	that takes arbitrary class hierarchies of programs written in object-oriented languages into account.
	The concept discusses interactions between constraint-logic object-oriented programs and reference type logic variables,
	particularly invocations on and access to logic variables, type operations, and equality.
	Furthermore, it proposes approaches as to how these interactions can be handled by a corresponding execution environment.

	\end{abstract}

\section[Motivation]{Motivation\footnote{This is a preprint of the work published in Functional and Constraint Logic Programming (WFLP 2018). The final authenticated version is available online at \url{https://doi.org/10.1007/978-3-030-16202-3_8}.}} \label{sec:intro}

Constraint-logic object-oriented programming can be used to develop 
business software that involves finding solutions to constraint-logic problems in an integrated way, particularly 
for applications that add constraints dynamically during runtime.
The mixed paradigm leverages benefits of well-known object-oriented programming languages as well as of constraint-logic programming.
For example, the constraint-logic object-oriented programming language Muli augments Java with logic variables, symbolic execution, constraints, and encapsulated search using a customised symbolic Java virtual machine (SJVM) \cite{Dageforde2018}. 

So far, symbolic expressions in Muli can involve logic variables of any type, but constraints can only be defined over (logic) variables of \textit{primitive} types~\cite{Dageforde2018semantics}. 
While those variables may be fields of objects, thus proving useful in an imperative context as well as in an object-oriented one, 
such constraints are not applicable to entire objects. 
Similarly, the semantics of further interactions (particularly invocations and field accesses) with unbound reference type logic variables is not defined yet.
After all, objects in object-oriented languages usually do not just encapsulate data, but behaviour as well.
As a result, such interactions lead to interesting behaviour, e.\,g., when methods are invoked on unbound logic variables or objects are compared for equality.
In order to realise the benefits of an integrated programming language,
the expected behaviour of such interactions needs to be defined and implemented.

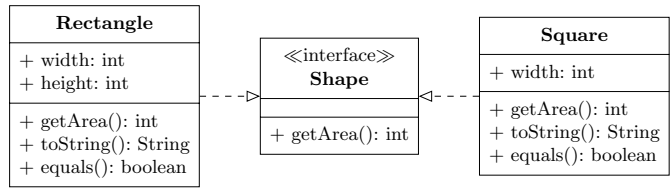
\begin{figure}
	\centering
	\scalebox{0.8}{
		\begin{tikzpicture}
		\tikzumlset{fill class=white}
		\umlinterface{Shape}{}{+ getArea(): int}
		\umlclass[right=1 cm of Shape.east]{Square}{+ width: int}{+ getArea(): int\\+ toString(): String\\+ equals(): boolean}
		\umlclass[left=1 cm of Shape.west]{Rectangle}{+ width: int\\+ height: int}{+ getArea(): int\\+ toString(): String\\+ equals(): boolean}
		\umlimpl{Square}{Shape}
		\umlimpl{Rectangle}{Shape}
		\end{tikzpicture}
	}
	\caption{Class structure assumed for the running example.}
	\label{fig:shapes}
\end{figure}

Consider the following case that will be used as a running example. We have an object-oriented representation of shapes, namely \lstinline|Rectangle| and \lstinline|Square| that both implement an interface \lstinline|Shape| (cf.~\Cref{fig:shapes}), assuming integer edge lengths in millimetres. Implementations of \lstinline|Shape| provide an appropriate method \lstinline|getArea()| that calculates the area from field values of an object, as well as a method \lstinline|toString()| that outputs the object's field values in a human-readable form.%
\footnote{Even though \lstinline|toString()| is not declared explicitly in the given interface, the Java language specification implicitly augments interfaces with abstract methods that correspond to every method that is declared in \lstinline|java.lang.Object| \cite[\S\ 9.2]{jls8}. Among others, this includes an implicit declaration of \lstinline|toString()| that is consistent with the corresponding declaration in \lstinline|Object|.}

As a simple example, \Cref{lst:shape-area} formulates a constraint to search for arbitrary shapes that have an area of $16$ square millimetres.
No specific instance is provided for \lstinline|s|; instead, \lstinline|s| is declared as a logic variable. On invocation of either \lstinline|getArea()| or \lstinline|toString()| on \lstinline|s|, the execution environment has to consider that multiple implementations of these methods are applicable, as per the definitions depicted in \Cref{fig:shapes}.
In Muli, we expect the applicable alternatives to be evaluated non-deterministically until all alternatives are considered \cite{Dageforde2018} (``don't know'' non-determinism), here resulting in at least two output lines, namely one per actual type of \lstinline|s|. 
Among other things, this paper will elaborate and discuss where exactly non-determinism can be introduced during the evaluation of this example and similar programs.

\begin{lstlisting}[caption={A constraint-logic object-oriented program that involves a free object.},label={lst:shape-area}]
Shape s free;
if (s.getArea() == 16) { 
	System.out.println(s.toString()); }
else { Muli.fail(); }
\end{lstlisting}

This paper contributes a concept for reference type logic variables in the context of constraint-logic object-oriented programming.
To that end, all types of interactions of a program with reference type logic variables are discussed based on the example of Muli.
This takes peculiarities of comparing equality of Java objects into account.
For each possible interaction, this paper defines the expected behaviour and outlines
 approaches for handling it in the context of arbitrary object graphs.
These approaches account for varying positions of objects' types in the class hierarchy that result from inheritance and implementation relations between classes.

This paper presents the contribution as follows.
\Cref{ref:muli} provides a brief introduction to the constraint-logic object-oriented programming language Muli.
Afterwards, \Cref{ref:approaches} discusses interactions and explains how they can be handled. Furthermore, that section introduces constraints that are necessary to achieve these interactions.
As this is  a report on research in progress, \Cref{ref:implementation} presents an initial implementation idea for a prototype that is going to be used for evaluation.
Related research is outlined in \Cref{ref:relatedwork}.
Finally, \Cref{ref:conclusion} summarises the contribution and provides an outlook.

\section{Constraint-logic Object-oriented Programming with Muli} \label{ref:muli}

As a constraint-logic object-oriented language,
  Muli allows developers to use programming styles of object-oriented programming,
  while facilitating the specification of constraint-logic problems and finding solutions to them in the same language \cite{Dageforde2018}.
Muli syntax is based on Java 8.
The SJVM serves as the execution environment 
that supports logic variables by means of symbolic execution 
  and leverages a constraint solver to solve constraint-logic problems.
Compared to Java, the syntax extension is minimal and limited to the \lstinline|free| keyword. It occurs in declaration statements to indicate an unbound (``free'') variable:

\begin{lstlisting}[xleftmargin=\parindent]
int x free;
\end{lstlisting}

At runtime, free variables of primitive types are treated as logic variables to be used as part of symbolic expressions.
Similarly, free objects can be defined, but their semantics is undefined and the execution environment does not provide an implementation for treating such variables yet. Therefore, the following code compiles but invoking the method in the second line will fail:

\begin{lstlisting}[xleftmargin=\parindent]
Object o free;
o.toString();
\end{lstlisting}

All variables, including unbound ones, can be used in boolean or arithmetic expressions in the same way as in Java.
However, if an expression contains unbound variables, they cannot evaluate to a specific value. Therefore, the execution environment treats those variables symbolically and creates a symbolic expression \cite{Dageforde2018semantics}.
For instance, after executing \Cref{lst:symbolic}, \lstinline|y| holds the constant value \lstinline|5| (as expected in Java), whereas \lstinline|z| holds the symbolic expression \lstinline|x + 5|.

\begin{minipage}{\linewidth}
\begin{lstlisting}[caption={Arithmetic expressions containing bound or unbound variables.}, label={lst:symbolic}]
int x free;
int i = 2, j = 3;
int y = i + j; // y == 5.
int z = x + y; // z == x + 5.
\end{lstlisting}
\end{minipage}

Ultimately, symbolic arithmetic expressions can evaluate to numeric constants (e.\,g., after labelling symbolic variables they contain). Therefore, an arithmetic expression that contains only \lstinline|int| (logic) variables and \lstinline|int| constants can be used anywhere where an \lstinline|int| expression is expected.

The behaviour described so far is deterministic.
However, as soon as a symbolic expression is used as part of a condition that leads to branching (e.\,g., in an \lstinline|if| statement), it is possible that the execution environment cannot decide on a unique outcome, e.\,g. whether a condition evaluates to \lstinline|true| or \lstinline|false|. 
When there is more than one choice, non-determinism is introduced, so that execution may continue with any of the possible branches \cite{Dageforde2018semantics}.
The execution environment makes a choice by selecting a branch, thus asserting a particular outcome (e.\,g., the condition shall be \lstinline|false|). That assertion is maintained by imposing a corresponding constraint on the constraint store.
After executing that branch, the execution environment backtracks  state (constraint store, operand and frame stacks, program counter, and heap values) to the point where a choice was made, and then proceeds with the next choice.
In Muli, this behaviour is referred to as search.

In order to limit the effects of non-deterministic execution, non-deterministic branching has to be encapsulated in the program. To that end, Muli offers encapsulation methods such as \lstinline|getAllSolutions()| or \lstinline|getOneSolution()| that take a lambda expression or a method reference as a parameter which is then executed non-deterministically.
The result of non-deterministic branching is a symbolic execution tree \cite{King1976}. Solutions to a constraint-logic problem correspond to the leaves of that tree, i.\,e. where execution ends, such as by throwing an exception or returning a value or expression.
The encapsulation method collects the required solutions and returns them to the calling, deterministic program.

\section{Reference Type Logic Variables (or Free Objects)} \label{ref:approaches}

As Muli is based on Java, Muli distinguishes the same four kinds of reference types as Java~ \cite[\S\ 4.3]{jls8}: class types, interface types, array types, and type variables.
Type variables are fundamentally different from the other kinds,
as they are substituted by a reference type.
For example, \lstinline|ArrayList<E>| contains the type variable \lstinline|E| that is substituted by a reference type, e.\,g., \lstinline|Object| or \lstinline|String|.
In contrast, the other kinds of reference types imply that they are instantiated at runtime with values that come from the heap, i.\,e. they point to data structures such as objects or arrays.
Since type variables are that different,
they are excluded from further considerations in this work,
 resulting in a definition of reference types that is congruent to that of C\# \cite{Microsoft2015}.%
 \footnote{Note that only the standalone use of type variables is disregarded here. Consequently, the reference types that we consider in the following may still make use of type variables as part of parameterised (generic) types.}
Class and interface types exhibit an identical structure \cite{jls8},
whereas array types are interpreted differently.
Even though array types are interesting as well, this work focuses on class and interface types for now. In the following, they are subsumed as \textit{reference types} for improved legibility.

Due to the nature of Java (and, therefore, Muli),
the reference types that this work focuses on 
  are not limited to data encapsulation. They also encapsulate behaviour (via methods) that may change along the implementation hierarchy as a consequence of overriding.
Therefore, when a variable that is declared by \lstinline|Object o| is of type \lstinline|Object|, \lstinline|o| may hold an instance of \lstinline|Object| or of its subclasses. This affects the typecasts that can (validly) be performed on \lstinline|o| at runtime, as well as the behaviour that is expected from invoking methods on the object.
This implies that interactions with a reference type logic variable declared by \lstinline|Object o free| need to consider that \lstinline|o| may represent instances of subclasses of \lstinline|Object| as well.

Consequently, we first need to define at which point exactly
non-determinism may be introduced when interacting with reference type logic variables. 
Options are 
either during declaration/initialisation of a reference type logic variable (i.\,e. at \lstinline|Object o free|), or when a feature of a variable that is not sufficiently specified is required later during runtime (e.\,g., on invocation of \lstinline|o.toString()| or on access to a field such as \lstinline|square.width|).
If non-determinism were already introduced at declaration/initialisation time, this would introduce many branches that are potentially irrelevant, because the SJVM cannot determine how many choices will be required. 
Therefore, aiming to reduce the state space, Muli creates choices only if discriminating behaviour is expected, e.\,g., when control flow branches.
For reference type logic variables, discriminating behaviour is not expected at the declaration of a logic variable (which can be done deterministically) 
but can be expected when one of its fields is accessed
or its methods are invoked.
Hence, we propose that 
non-determinism is incurred when a feature of a logic variable $v$ is required, where $v$ is not sufficiently specified to be handled deterministically.
As a result, this
allows search to focus on branches relevant to the respective access, thus effectively reducing the state space.
Note that these considerations are similar to those regarding the \textit{Label} reduction rule from~\cite{Dageforde2018semantics} that is used for substituting primitive type logic variables with their potential values.
Similar to the present case, \textit{Label} is suggested to be used only as a last resort if no other rule can be applied
as its application results in one branch per potential value, which usually are a lot.
If this is done too early during evaluation, this increases the state space unnecessarily~\cite{Dageforde2018semantics}.

With this in mind, there are six different kinds of interactions between a program and a reference type logic variable that need to be examined in the following as they potentially result in non-determinism.
First, accesses to fields of an object by a program, followed by invocations of methods.
Moreover, the program can compare equality, which occurs in two forms in Java (and therefore in Muli), i.\,e. comparing reference equality or value equality, which are the third and fourth kind, respectively. 
Fifth, a program can perform operations on the type of a variable.
Last but not least, as a novel kind of interaction,
programmers may expect to be able to compare objects for structural equality, i.\,e. equality based on objects' field values instead of the entire object.
 This is similar to unification of constructor terms, which is common in logic programming but not in object-oriented programming languages.

\subsection{Accessing a field of a free object} \label{sec:accessfields}

In Muli and Java, fields are accessed using a dot notation, e.\,g., \lstinline|square.width|.
In contrast to methods, fields of a Java class cannot be overridden by subclasses. Although subclasses can declare fields with names identical to those in superclasses, this merely results in the original field being hidden from the overriding class, but not from the original one.
Consider an artificial Java example in \Cref{lst:fields}. Accesses to \lstinline|i| in both cases \lstinline|a.i| and \lstinline|b.i| result in the same value 2 because \lstinline|a| and  \lstinline|b| are accessed via variables of type \lstinline|A|. Of course, if \lstinline|b| were stored in a variable of type \lstinline|B|, that would not be the case. Muli shares this semantics with Java.

\begin{minipage}{\linewidth}
\begin{lstlisting}[label=lst:fields,caption={Fields are only hidden, but not overridden.}]
class Demo {
	public static void main(String[] args) {
		A a = new A();
		A b = new B(); } }
class A { public int i = 2; }
class B extends A { private int i = 1; }
\end{lstlisting}
\end{minipage}

As a result, accesses to fields of free objects do not need to consider the class hierarchy of the object's type, but only the type of the reference type logic variable through which access takes place (here, \lstinline|A|).
Since a free object is uninitialised, in its initial state all its fields are to be treated as logic variables as well.
Therefore, accessing a field of a free object is a deterministic operation.
Its result is the logic variable that is the field of the object. For instance, in the running example accessing \lstinline|square.width| yields the logic variable of type \lstinline|int| that is stored at that field.

\subsection{Invoking a method on a free object} \label{sec:invokemethod}

For a variable \lstinline|Shape s free|,
consider the statement \lstinline|s.getArea()| as seen in \Cref{lst:shape-area}. 
As \lstinline|s| is declared free, this causes the execution to evaluate the method \lstinline|getArea()|.
\lstinline|Shape| is merely an abstract supertype,
so all the subtypes need to be taken into consideration, as they provide implementations for \lstinline|getArea()|.
Similarly, even in the deterministic nature of Java, the method that is actually invoked depends on the type of the referenced instance, not on that of the variable.
Consequently, in order to determine which actual implementation is going to be invoked, the statement \lstinline|s.getArea()| causes the SJVM to discover the set $S$ of non-abstract subtypes that extend \lstinline|Shape|.%
\footnote{In general, this includes parameterised (generic) types that remain in parameterised form (e.\,g., \lstinline|ArrayList<E>|). Therefore, this set is finite.} 
If the supertype can be instantiated as well, 
the set of relevant types then is $S' = S \cup \{\mathtt{X}\}$ for a non-abstract supertype \lstinline|X|. Otherwise, the set of relevant types is just $S' = S$. 
For the running example, $S' = S = \{\mathtt{Square}, \mathtt{Rectangle}\}$, as the supertype is an interface type and is therefore abstract.

In general, the set of relevant subtypes can be restricted further, thus reducing the number of non-deterministic branches that the SJVM needs to evaluate.
After all, we are only interested in those branches that potentially exhibit distinct behaviour.
Therefore, the SJVM needs to discover $S'' \subseteq S'$, comprising only those classes that provide their own implementations of \lstinline|getArea()|, thus omitting all types that merely inherit an implementation from their supertype.
Afterwards, the SJVM only needs to evaluate one branch per element of $S''$.
If $S''$ holds exactly one type, execution continues deterministically by invoking that type's implementation on $s$. Otherwise, evaluation creates a choice point in order to execute all \lstinline|((t)s).getArea()|, where $\mathtt{t} \in S''$.
As a result, the number of choices that this choice point provides is equal to the cardinality of $S''$.

\begin{figure}
	\centering
	\scalebox{0.8}{
		\begin{tikzpicture}
		\tikzumlset{fill class=white}
		\umlclass{A}{}{+ m()}
		\umlclass[right=1cm of A.east]{B}{}{+ m()}
		\umlclass[right=1cm of B.east]{C}{}{}
		\umlclass[right=1cm of C.east]{D}{}{+ m()}
		\umlinherit{B}{A}
		\umlinherit{C}{B}
		\umlinherit{D}{C}
		\node[draw=ercisred,thick,inner xsep=10pt,fit={(B)(C)}](appl){};
		\node[anchor=east,below left=0.3cm and 0.3cm of B.west |- appl.south](l1){selected implementation of \lstinline|m()|};
		\node[anchor=west, below right=0.3cm and -0.1cm of C.east |- appl.south, ercisred](l2){applicable instance types after choosing \textbf{B}};
		\node[anchor=west, above right=0.2cm and 2.3cm of C.east |- appl.north](l3){applicable instance types};
		\draw[->, ercisred] (l2) -| (C |- appl.south);
		\draw[->, shorten >=-8pt] (l1) -| (B |- appl.south);
		\draw[->] (l3) -| (A.north);
		\draw[->] (l3) -| (B.north);
		\draw[->] (l3) -| (C.north);
		\draw[->] (l3) -| (D.north);
		\node[below=0.1cm of l3.south east,anchor=east](l4){= $S''$};
		
		\end{tikzpicture}
	}
	\caption{Applicable instance types for a given object \lstinline|A a free| before and after choosing a particular subtype.}
	\label{fig:applicabletypes-method}
\end{figure}
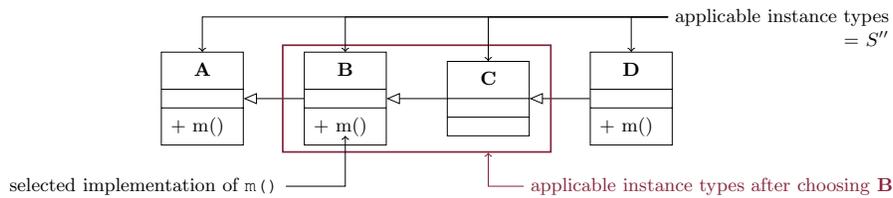

Looking at the running example from  \Cref{lst:shape-area}, $S'$ cannot be reduced as all subtypes provide their own implementations, i.\,e. $S'' = S' = \{\mathtt{Square}, \mathtt{Rectangle}\}$.
For this reason the \lstinline|System.out.println| statement is expected to be executed twice, as indicated in \Cref{sec:intro}; once per type in $S''$.
To discuss a different example with a more detailed implementation hierarchy, consider the classes depicted in \Cref{fig:applicabletypes-method}.
For a logic variable \lstinline|A a free|, invoking \lstinline|a.m()| results in discovering the subtypes $S = \{\mathtt{B}, \mathtt{C}, \mathtt{D}\}$ first. 
The supertype \lstinline|A| is non-abstract, therefore $S' = \{\mathtt{A}, \mathtt{B}, \mathtt{C}, \mathtt{D}\}$. However, since \lstinline|C| does not provide its own implementation of \lstinline|m()| and relies on that of \lstinline|B| instead, the set is reduced further to 
 $S'' = \{\mathtt{A}, \mathtt{B}, \mathtt{D}\}$. The SJVM then continues the evaluation based on $S''$.

After making a choice for a type $t \in S''$
whose method implementation is used, the actual type of the instance that the method is invoked on can be an arbitrary one from a set of types. Specifically, either the determined type or any of its subtypes. However, the set of allowed types is restricted further, as it may not contain subtypes that provide their own implementation (as \textit{their} implementation would need to be invoked otherwise).
This is illustrated in \Cref{fig:applicabletypes-method}, where the set of types is constrained only to \lstinline|B| and \lstinline|C|. Even though \lstinline|D| is a subtype, it provides an own implementation of \lstinline|m()| and would therefore conflict with having chosen \lstinline|B|'s implementation.

As a result of choosing an implementation, the SJVM needs to add a constraint to its constraint store that restricts the type of \lstinline|s| according to the above description.
This ensures that later interactions with that object do not make conflicting assumptions regarding the type of \lstinline|s|, i.\,e. to avoid assuming \lstinline|s| to be of a type that is not in the reduced set of applicable types.
Similarly, a type $t$ cannot be assumed for \lstinline|s| if that would violate a previously imposed constraint, so the corresponding branch must not be evaluated. 
Consequently, the constraint that restricts an instance's type is a set-based constraint. This type of constraint is novel to Muli, as existing constraints are only of arithmetic nature.

\subsection{Comparing reference equality of reference type logic variables} 
\label{sec:referenceeq}

In Muli and Java, objects are typically compared by one of two means, either reference equality or value equality. First, let us focus on the former.
Based on the program in \Cref{lst:genesis},
consider the conditional control flow statements \lstinline|if (o == p)| and \lstinline|if (o == q)| that compare references of reference type (logic) variables.

\begin{lstlisting}[caption={Declaration of a set of reference type variables.}, label={lst:genesis}]
Object o free;
Object p = new Object();
Object q free;
\end{lstlisting}

As \lstinline|o| and \lstinline|q| are declared free, it needs to be discussed whether the constraint created by evaluating reference equality  
should result in the SJVM unifying their references upon evaluation of the condition, i.\,e. result in \lstinline|o| pointing to the instance referenced by \lstinline|p| (for \lstinline|o == p|), or to the same reference as the other logic variable \lstinline|q| (for \lstinline|o == q|).
Arguably, this should not be the case. \Cref{lst:genesis} expressly declares the three variables to be three different instances, unlike an assignment, such as \lstinline|Object w = p|, which would explicitly make \lstinline|w| assume the same reference as \lstinline|p|.

Therefore, the evaluation of a condition comparing reference equality is a deterministic operation even for reference type logic variables that yields \lstinline|true| iff two variables reference the same free object, which is consistent with the Java semantics of comparing reference equality. No implicit unification is performed.

\subsection{Comparing value equality of reference type logic variables} 
\label{sec:valueeq}

In addition to the means described in \Cref{sec:referenceeq},
Java (or Muli) code can also compare objects
in terms of value equality, e.\,g., by \lstinline|if (o.equals(p))| or \lstinline|if (p.equals(o))| (after an initialisation as depicted in \Cref{lst:genesis}).
This presents another opportunity for unifying objects if free objects are involved.

As \lstinline|equals()| is a method that every class can implement individually, the interpretations of these two examples are fundamentally different.
In \lstinline|p.equals(o)|, \lstinline|p| is a concrete instance of \lstinline|Object|, so \lstinline|Object|'s default implementation is invoked deterministically, 
effectively checking for reference equality. Other implementations might compare instances by accessing fields of the free object \lstinline|o|, thus resorting to the case described in \Cref{sec:accessfields}.
In contrast, \lstinline|o.equals(p)| is an invocation of \lstinline|equals()| on the logic variable \lstinline|o|. As a result, this case reduces to the invocation of methods (cf.\ \Cref{sec:invokemethod}), resorting to specific implementations of \lstinline|equals()|, e.\,g., of \lstinline|Square| and \lstinline|Rectangle|.
Consequently, \lstinline|equals()| is not commutative.

As a result, Muli does not need to handle value equality comparisons specifically, as they are implicitly covered by other considerations regarding reference type logic variables.

\subsection{Performing type operations on a free object} \label{sec:typecasts}
The (super-) type of a logic variable is determined by its declaration, but initially the corresponding instance may be of that type or of its subtypes (cf. the definition of $S$ in \Cref{sec:invokemethod}).
This affects operations that operate on the type of a free object; namely \lstinline|instanceof| and typecasts. 
For example, the set of allowed types for the instance is reduced by (successful or failed) typecasts.
Considering \Cref{lst:genesis} again, a program might try to cast a reference type logic variable to a subtype, e.\,g., \lstinline|(Square)o|.
In that case, given that this is a valid cast, the actual type of \lstinline|o| can be  \lstinline|Square| or any of its subtypes.

Typecasts can be either valid or invalid at runtime.
Invalid typecasts are those that violate the class hierarchy, such as casting an object of type \lstinline|Square| to \lstinline|Rectangle|.
This deterministically yields a \lstinline|ClassCastException|
and therefore does not result in a choice point.
The result of evaluating \lstinline|instanceof| statements in a similarly invalid contexts is deterministically \lstinline|false|.

In contrast, 
performing a valid typecast results in two choices as to how execution can continue.
Either the cast is successful (unless a contradictory constraint exists in the constraint store at runtime), so a new constraint can be imposed narrowing the logic variable's type;
or the cast is not successful. In regular Java, the latter case is not caught by a compiler and results in a runtime exception (\lstinline|ClassCastException|). Similarly, Muli can handle this case by imposing a corresponding constraint and throwing that exception.
Therefore, a valid typecast of a reference type logic variable results in a choice point with at most two options, depending on existing constraints in the constraint store.
Similarly, using \lstinline|instanceof| in a valid context results in non-deterministic execution that imposes the same constraints as successful or unsuccessful typecasts.

To support non-deterministic branching, a constraint is needed that is imposed when a choice is made for a branch that corresponds to a type operation.
This constraint reduces the set of possible instance types.
The set-based constraint from \Cref{sec:invokemethod} can be re-used, but the sets are computed differently.
Given that $S$ describes the set of applicable types prior to imposing a constraint and $U$ describes the set of types comprising the cast target types and all of its subtypes,
on a successful cast, the set of applicable types is narrowed to the intersection $V = S \cap U$,
 whereas for a failed typecast all remaining types are applicable, i.\,e. the type is constrained to the set difference $V' = S \setminus U$.
The resulting sets of types are used to impose the corresponding constraints, i.\,e. $V$ for the constraint that is added to the constraint store when making the choice that the typecast is successful, and $V'$ for the other choice.

\subsection{Imposing a constraint for structural equality between two objects} \label{sec:structuralequality}

The cases discussed so far refer to the interpretation of object-oriented concepts against the background of a constraint-logic object-oriented language.
In addition to that, Muli creates a novel opportunity regarding unification of objects that cannot exist in plain object-oriented languages without symbolic execution, namely
comparing (free) objects for structural equality (in combination with constraints that enforce it).

Value equality relies on the \lstinline|equals()| method that a class can implement individually (cf.\ \Cref{sec:valueeq}), for example so that equality depends only on a specific field.
In contrast, 
we use the term structural equality to refer to a situation in which all fields of two (free) objects of the same type either share identical values (for fields of primitive types) or are structurally equal again (for reference-type fields), i.\,e. the following recursive definition applies:
$ o_1 \odot o_2 \Leftrightarrow type(o_1) = type(o_2) \land (
	(o_1.x\ \mathrm{primitive} \land o_1.x = o_2.x)
	\lor
	(o_1.x\ \mathrm{not\ primitive} \land o_1.x \odot o_2.x)
)\ \forall x \in fields(o_1)$,%
\footnote{Note that here $fields(o_1) = fields(o_2)$ since $type(o_1) = type(o_2)$, so $fields(o_2)$ could be used just as well.}
 where $type(o)$ is the type of an object $o$ and $fields(o)$ is the set of its fields.
For example, given two free objects \lstinline|Rectangle r1 free, r2 free|, imposing structural equality \lstinline|r1| $\odot$ \lstinline|r2| implies that \lstinline|r1.width == r2.width| and \lstinline|r1.height == r2.height| in addition to sharing their type. Similarly, if \lstinline|r2| were an initialised object of type \lstinline|Rectangle|, the values of \lstinline|r1|'s fields are unified with those of the corresponding fields in \lstinline|r2|.
As a result, 
  \lstinline|r1| $\odot$ \lstinline|r2| $\Leftrightarrow$
  \lstinline|r2| $\odot$ \lstinline|r1|, i.\,e. structural equality is commutative.

A new operator is needed to denote the structural equality constraint  $\odot$ in source code.
For that purpose, I introduce the symbol \lstinline|#=| to be used as a boolean, binary operator in conditions
  in order to add this constraint to the constraint store at runtime.
It evaluates to \lstinline|true| if fields of two objects are unifiable as described above, and to \lstinline|false| if they are not. In both cases, a corresponding constraint is added to the constraint store that maintains this equality.

\section{Implementation} \label{ref:implementation}

The considerations in \Cref{ref:approaches} require modifications to the Muli SJVM in terms of additional constraints and choice point types. This results in changes that need to be made to the SJVM's solver component and its choice point generator (cf. \cite{Dageforde2018}).

The \textit{applicable type constraint} is a set-based constraint that restricts possible types for a free object.
It maintains a reference to the free object that it affects, and a set of fully qualified names of types that the object may assume.
This set is defined prior to instantiation of that constraint.
In the solver manager, a constraint is imposed in conjunction with all other constraints in the constraint store. Therefore,
the solver manager can verify consistency of a constraint store by collecting all imposed applicable type constraints involving a free object and checking that the intersection of the sets of types is non-empty for each object, i.\,e. there is at least one type that any object can assume; in addition to verifying consistency of the remaining constraints.

Additionally, the \textit{structural equality constraint} translates into a conjunction of arithmetic equality and type equality constraints as specified in \Cref{sec:structuralequality}, hence it does not need to be represented on its own.
The \textit{type equality constraint} references two involved objects that need to be of the same type.
A constraint store comprising a type equality constraint is consistent if both objects are trivially of the same type (such as for regular objects) or if there is a type that is among the applicable types of both objects.

At runtime, evaluations of bytecode instructions that incur non-determinism result in the creation of choice points. These are responsible for controlling search and, hence, for imposing constraints and removing them afterwards \cite{Dageforde2018}.
Therefore, the support for type operations on logic reference type variables requires a
corresponding choice point.
It offers choices according to the description in \Cref{sec:typecasts} and imposes an appropriate instance of the applicable type constraint for each  choice.
Similarly, a choice point for invoking a method according to \Cref{sec:invokemethod} is required.
Both choice point implementations require the implementation of new helper methods that discover sets of available types.
The method \lstinline|Type[] getSubtypes(Type)| discovers, for a given type, all of its subtypes from the loaded classpath.
A further method \lstinline|Type[] getImplementations|\allowbreak\lstinline|(Type[], Method)| is required that filters a list of types such that it returns only those types that can be instantiated and that provide an own implementation of a particular method, thus supporting the case from \Cref{sec:invokemethod}.

Last but not least, another choice point is generated if free objects are compared for structural equality  as specified in \Cref{sec:structuralequality}. It comprises two choices. One choice represents that equality is maintained, resulting in the corresponding constraint being imposed.
The other one corresponds to imposing the negation of that constraint.

\section{Related Work} \label{ref:relatedwork}

Several approaches intend to integrate elements from object-oriented programming into declarative languages, mostly based on Prolog.
For example, tuProlog provides a Prolog engine implemented in Java, offering access to Java features from Prolog \cite{Denti2005}. However, referring to Java types is done rather artificially by means of string literals which cannot be checked by a compiler, and free objects and accessing their fields are not considered.
As a non-Prolog-based example, Oz is a constraint language that offers OO features, but does not seem to support constraints involving logic objects \cite{VanRoy2003}.
Despite their integration, the mentioned programming languages follow a declarative style, which might not be as accessible for developers who are used to imperative languages.

CAPJa intends to seamlessly integrate Prolog search into Java programs, e.\,g. by providing a Java-based abstraction layer from Prolog \cite{DBLP:conf/ruleml/Ostermayer15}. The integration supports a mapping of data structures from Java to Prolog and vice-versa, but focuses on logic objects used for encapsulating data. It does not consider free (unbound) objects in terms of method invocations and field accesses, which become relevant if we consider that objects also encapsulate behaviour, which is expected in object-oriented programming.
As another example, 
the library \texttt{heya-unify} facilitates unification of data structures in JavaScript \cite{Lazutkin2014}, particularly in order to compare object contents or to perform pattern matching on them. However, it does not support defining entire objects as logic variables and is limited to comparing structural equality on weakly typed objects and arrays.

The type unification algorithm presented by \cite{Plumicke2009} can be used for Java type inference.
Although their work emanates from a different standpoint, the type unification could be re-used for formulating the subtype relations for the constraints in this work.

Other work demonstrates that the use of languages integrating multiple paradigms is beneficial, most notably the Java Stream API \cite{Urma2014} and  Scala \cite{scala}, which integrate object-oriented programming with functional programming on the JVM.
LINQ offers a similar integration, but for languages on the .NET CLR~\cite{Meijer2006}.
A very relevant integration of logic and functional programming is Curry \cite{Hanus1995}, which incorporates logic programming into a language with Haskell syntax. 
Muli lends and adapts some ideas from Curry, such as encapsulated search and constraint definition via boolean equalities \cite{Antoy2015}. However, the adaptation of these concepts to constraint-logic \textit{object-oriented} programming results in fundamentally different considerations and implementations.

\section{Concluding Remarks} \label{ref:conclusion}

This work contributes a concept for reference type logic variables in constraint-logic object-oriented languages. 
It details interactions of programs with reference type logic variables and discusses approaches for handling such interactions,
on the basis of the programming language Muli.
As a result, there now is a concept for 
  invocations on free objects and accesses to their fields,
  comparisons of different kinds of equality,
  and type operations
in constraint-logic object-oriented programming.

The discussed approaches efficiently introduce non-determinism where it is specifically required and take class hierarchies into account.
This requires a novel constraint that restricts types of free objects to support these approaches.
Since the constraints previously supported by Muli were of a purely arithmetic nature,
this work also contributes a set-based constraint to restrict the possible types of free objects.

The contribution is helpful not just for Muli but for constraint-logic object-oriented programming in general,
because it allows non-deterministic search to extend beyond logic variables of primitive types.
For example, a constraint-logic object-oriented language based on C\# could also make use of these approaches.
Furthermore, it 
facilitates the usage of object-oriented features in combination with free objects.

Subsequently, the implementation of this approach in the Muli SJVM will be completed in order to evaluate its benefits. The resulting virtual machine implementation will be part of the open source distribution of Muli provided via GitHub.\footnote{\url{https://github.com/wwu-pi/muli}.}
It is also planned to provide an augmented formal semantics, incorporating the aspects discussed in this paper, thus yielding an integrated semantics for a constraint-logic \textit{object-oriented} language.
Future work will tackle the extension of these considerations towards further reference types, particularly array types.

%
%
%
\bibliographystyle{splncs04}
\bibliography{lit}
	
\end{document}